\documentstyle{article}

\def\bea{\begin{eqnarray}}
\def\eea{\end{eqnarray}}
\def\nn{\nonumber}

\def\beq{\begin{equation}}
\def\eeq{\end{equation}}
\def\ba{\beq\new\begin{array}{c}}
\def\ea{\end{array}\eeq}
\def\be{\ba}
\def\ee{\ea}
\def\stackreb#1#2{\mathrel{\mathop{#2}\limits_{#1}}}

\makeatletter
\newdimen\normalarrayskip              
\newdimen\minarrayskip                 
\normalarrayskip\baselineskip
\minarrayskip\jot
\newif\ifold             \oldtrue            \def\new{\oldfalse}
\def\arraymode{\ifold\relax\else\displaystyle\fi} 
\def\eqnumphantom{\phantom{(\theequation)}}     
\def\@arrayskip{\ifold\baselineskip\z@\lineskip\z@
     \else
     \baselineskip\minarrayskip\lineskip2\minarrayskip\fi}
\def\@arrayclassz{\ifcase \@lastchclass \@acolampacol \or
\@ampacol \or \or \or \@addamp \or
   \@acolampacol \or \@firstampfalse \@acol \fi
\edef\@preamble{\@preamble
  \ifcase \@chnum
     \hfil$\relax\arraymode\@sharp$\hfil
     \or $\relax\arraymode\@sharp$\hfil
     \or \hfil$\relax\arraymode\@sharp$\fi}}
\def\@array[#1]#2{\setbox\@arstrutbox=\hbox{\vrule
     height\arraystretch \ht\strutbox
     depth\arraystretch \dp\strutbox
     width\z@}\@mkpream{#2}\edef\@preamble{\halign
\noexpand\@halignto
\bgroup \tabskip\z@ \@arstrut \@preamble \tabskip\z@ \cr}%
\let\@startpbox\@@startpbox \let\@endpbox\@@endpbox
  \if #1t\vtop \else \if#1b\vbox \else \vcenter \fi\fi
  \bgroup \let\par\relax
  \let\@sharp##\let\protect\relax
  \@arrayskip\@preamble}
\def\eqnarray{\stepcounter{equation}%
              \let\@currentlabel=\theequation
              \global\@eqnswtrue
              \global\@eqcnt\z@
              \tabskip\@centering
              \let\\=\@eqncr
              $$%
 \halign to \displaywidth\bgroup
    \eqnumphantom\@eqnsel\hskip\@centering
    $\displaystyle \tabskip\z@ {##}$%
    \global\@eqcnt\@ne \hskip 2\arraycolsep
         $\displaystyle\arraymode{##}$\hfil
    \global\@eqcnt\tw@ \hskip 2\arraycolsep
         $\displaystyle\tabskip\z@{##}$\hfil
         \tabskip\@centering
    &{##}\tabskip\z@\cr}
\newfont{\hr}{msbm10}
\newfont{\ams}{msam10}

\begin{document}
\begin{titlepage}
\setcounter{footnote}0
\begin{center}
\hfill FIAN/TD-10/96\\
\hfill ITEP/TH-22/96\\
\hfill hep-th/9607109\\
\vspace{0.3in}
{\LARGE\bf  WDVV-like equations in ${\cal N}=2$ SUSY Yang-Mills Theory}
\\
\bigskip\bigskip\bigskip

{\Large A.Marshakov
\footnote{E-mail address:
mars@lpi.ac.ru, andrei@rhea.teorfys.uu.se, marshakov@nbivax.nbi.dk}$^{\ddag}$,
A.Mironov
\footnote{E-mail address:
mironov@lpi.ac.ru, mironov@grotte.teorfys.uu.se}$^{\ddag}$,
A.Morozov
\footnote{E-mail address:
morozov@vxdesy.desy.de}
$^{\dag}$}\\
\bigskip
$\phantom{gh}^{\dag}${\it ITEP, Moscow ~117 259, Russia}\\
$\phantom{gh}^{\ddag}${\it Theory Department, P. N. Lebedev Physics
Institute, Leninsky prospect 53, Moscow, ~117924, Russia\\
and ITEP, Moscow ~117259, Russia}\\
\end{center}
\bigskip \bigskip

\begin{abstract}
The prepotential $F(a_i)$, defining the low-energy
effective action of the $SU(N)$ ${\cal N}=2$ SUSY gluodynamics,
satisfies an
enlarged set of the WDVV-like equations $F_iF_k^{-1}F_j = F_jF_k^{-1}F_i$ for
any triple $i,j,k = 1,\ldots,N-1$, where matrix $F_i$ is equal to $(F_i)_{mn}
= {\partial^3 F\over\partial a_i\partial a_m \partial a_n}$.  The same
equations are actually true for generic topological theories. In contrast to
the conventional formulation, when $k$ is restricted to $k=0$, in the
proposed system there is no distinguished ``first'' time-variable, and the
indices can be raised with the help of any ``metric'' $\eta_{mn}^{(k)} =
(F_k)_{mn}$, not obligatory flat. All the equations (for all $i,j,k$) are
true simultaneously.  This result provides a new parallel between the
Seiberg-Witten theory of low-energy gauge models in $4d$ and topological
theories.  \end{abstract}

\end{titlepage}

\newpage
\setcounter{footnote}0

\section{Definitions}

According to \cite{1}, \cite{SW} the low-energy effective
action of ${\cal N}=2$ SUSY Yang-Mills model (the Seiberg-Witten
effective theory) is given by
\be\label{1}
\int d^4x d^4\theta F(\Phi_i),
\ee
where the superfield $\Phi_i = \varphi^i + \theta\sigma_{\mu\nu}
\tilde\theta G_{\mu\nu}^i + \ldots$.

The prepotential $F$ \cite{SW} is defined in terms of a
family of Riemann surfaces,
endowed with the meromorphic differential
$dS$.  For the gauge group $G=SU(N)$ the family is
\cite{SW}, \cite{2}, \cite{GKMMM}
\be
w + \frac{1}{w} = 2P_N(\lambda ), \\
P_N(\lambda ) = \lambda ^N + \sum_{k=1}^{N-1} h_k\lambda ^{k-1},
\label{hesur}
\ee
and
\be
dS = \lambda\frac{dw}{w}
\label{dS}
\ee
The prepotential $F(a_i)$ is implicitly defined by the set of equations:
\be
\frac{\partial F}{\partial a_i} = a_i^D, \\
a_i = \oint_{A_i} dS, \\
a^D_i = \oint_{B_i} dS.
\label{defF}
\ee
According to \cite{GKMMM}, this definition identifies
$F(a_i)$ as logarithm of (truncated) $\tau$-function of
Whitham integrable hierarchy.
Existing experience with Whitham hierarchies \cite{KDM}
implies that $F(a_i)$ should satisfy some sort
of the Witten-\-Dijkgraaf-\-Verlinde-\-Verlinde (WDVV) equations
\cite{WDVV}.

\section{The statement}

Below in this paper we demonstrate that WDVV equations
for the prepotential actually look like
\be
F_i F_k^{-1} F_j = F_j F_k^{-1} F_i
\ \ \ \ \ \ \forall i,j,k = 1,\ldots,N-1.
\label{FFF}
\ee
Here $F_i$ denotes the matrix
\be
(F_i)_{mn} = \frac{\partial^3 F}{\partial a_i
\partial a_m\partial a_n}.
\ee

\section{Comments}

{\bf 3.1} Let us remind, first, that the conventional WDVV equations
for topological field theory express the associativity of the algebra
$\phi_i\phi_j = C_{ij}^k\phi_k$ (for symmetric in $i$ and $j$ structure
constants):
$(\phi_i\phi_j)\phi_k = \phi_i(\phi_j\phi_k)$,
or $C_iC_j = C_jC_i$, for the matrix  $(C_i)^m_n \equiv C_{in}^m$.
These conditions become highly non-trivial since, in
topological theory, the structure constants are expressed
in terms of a single prepotential $F(t_i)$:
$C_{ij}^l = (\eta^{-1}_{(0)})^{kl}F_{ijk}$, and
$F_{ijk} = {\partial^3F\over\partial t_i\partial t_j\partial t_k}$,
while the metric is $\eta_{kl}^{(0)} = F_{0kl}$, where $\phi_0 = I$
is the unity operator. In other words, the conventional WDVV equations
can be written as
\be
F_i F_0^{-1} F_j = F_j F_0^{-1} F_i.
\label{FFFconv}
\ee
In contrast to (\ref{FFF}), $k$ is restricted to
$k=0$, associated with the distinguished unity operator.

On the other hand, in the Seiberg-Witten theory there does not clearly exist
any distinguished index $i$: all the arguments $a_i$ of the
prepotential are on equal footing. Thus, if some kind of
the WDVV equations holds in this case, it should be invariant under
any permutation of indices $i,j,k$ -- criterium satisfied by
the system (\ref{FFF}).

Moreover, the same set of equations (\ref{FFF}) is satisfied for
generic topological theory: see s.4.1 below.

{\bf 3.2} In the general theory of Whitham hierarchies
\cite{KDM} the WDVV equations arise also in the form (\ref{FFFconv}).
Again, there exists  a distinguished time-variable
$t_0 = x$ -- associated with the first time-variable of the original
KP/KdV hierarchy. Moreover, usually -- in contrast to the
simplest topological models -- the set of these variables for
the Whitham hierarchy is infinitely large. In this context
our eqs.(\ref{FFF}) state that, for peculiar subhierarchies
(in the Seiberg-Witten gluodynamics, it is the Toda-chain hierarchy,
associated with a peculiar set of hyperelliptic surfaces),
there exists a non-trivial {\it truncation} of the
quasiclassical $\tau$-function, when it depends on the finite
number ($N-1=g=$ genus of the Riemann surface) of
{\it equivalent} arguments $a_i$, and satisfies a much wider
set of WDVV-like equations: the whole set (\ref{FFF}).

{\bf 3.3} From (\ref{defF}) it is clear that $a_i$'s
are defined modulo linear transformations
(one can change $A$-cycle for any linear combination
of them). Eqs.(\ref{FFF}) possess adequate ``covariance'':
the least trivial part is that $F_k$ can be
substituted by $F_k + \sum_l\epsilon_l F_l$. Then
$$F_k^{-1} \rightarrow (F_k + \sum \epsilon_lF_l)^{-1}
= F_k^{-1} - \sum\epsilon_l F_k^{-1}F_lF_k^{-1} +
\sum \epsilon_l\epsilon_{l'} F_k^{-1}F_lF_k^{-1}F_{l'}
F_k^{-1} + \ldots$$ Clearly, (\ref{FFF}) -- valid for all
triples of indices {\it simultaneously} -- is enough to guarantee that
$F_i(F_k + \sum \epsilon_lF_l)^{-1}F_j =
F_j(F_k + \sum \epsilon_lF_l)^{-1}F_i$.
Covariance under any replacement of $A$ and $B$-cycles together will be
seen from the general proof in s.4 below: in fact the role
of $F_k$ can be played by $F_{d\omega}$, associated with
{\it any} holomorphic 1-differential $d\omega$ on the
Riemann surface.

{\bf 3.4} For metric $\eta$, which is a second derivative,
\be
\eta_{ij} = \frac{\partial^2 h}{\partial a_i\partial a_j}
\equiv h_{,ij}
\ee
(as is the case for our
$\eta^{(k)}_{mn}\equiv (F_k)_{mn}$: $\ h = h^{(k)} = \partial
F/\partial a_k$),
$\Gamma^{i}_{jk} = \frac{1}{2}\eta^{im}h_{,jkm}$ and the
Riemann tensor
\be
R^i_{jkl} =
\Gamma^i_{jl,k} + \Gamma^i_{kn}\Gamma^n_{jl} - (k \leftrightarrow l) =
\frac{1}{2}\eta^{im}h_{,jklm} -{1\over 4}
\eta^{ip}h_{,pnk}\eta^{nm}h_{,mjl}
 - (k\leftrightarrow l) =
\nn \\=
- \Gamma^i_{kn}\Gamma^n_{jl} + (k \leftrightarrow l)
= -{1\over 4}
\eta^{ip}h_{,pnk}\eta^{nm}h_{,mjl}
+ (k\leftrightarrow l)
\ee
In terms of the matrix $\eta = \{(\eta)_{kl}\}$ the
zero-curvature condition $R_{ijkl} = 0$ would be
\be
\eta_{,i}\eta^{-1}\eta_{,j} \stackrel{?}{=}
\eta_{,j}\eta^{-1}\eta_{,i}.
\label{R}
\ee
This equation is remarkably similar to (\ref{FFF}) and
(\ref{FFFconv}), but when $\eta^{(k)}_{ij} = F_{ijk}$ is
substituted into (\ref{R}), it contains the {\it fourth}
derivatives of $F$:
\be\label{FFFD}
F_{k,i}F^{-1}_{k}F_{k,j} \stackrel{?}{=} F_{k,j}F^{-1}_{k}F_{i,k}
\ \ \ \ \ \forall i,j,k = 1,\ldots,N-1
\ee
(no summation over $k$ in this formula!), while (\ref{FFF})
is expressed through the third derivatives only.

In ordinary topological theories $\eta^{(0)}$ is always flat,
i.e. (\ref{FFFD}) holds for $k=0$ along with (\ref{FFFconv}) -
and this allows one to choose ``flat coordinates'' where
$\eta^{(0)} = const$. Sometimes - see Appendix B for an
interesting example - {\it all} the metrics $\eta^{(k)}$
are flat simultaneously. However, explicit example of s.5.1.2
demonstrates that this is not always the case: in this
example (quantum cohomologies of $CP^2$) eqs.(\ref{FFF})
are true for all $k=0,1,2$, but only $\eta^{(0)}$
is flat (satisfies (\ref{R})), while $\eta^{(1)}$ and
$\eta^{(2)}$ lead to non-vanishing curvatures.

{\bf 3.5} Throughout this paper we do not include $\Lambda_{QCD}$
(the remnant of the dilaton v.e.v.) in the set of moduli.
Thus, our prepotential is a function of $a_i$ alone and does
not need to be a homogeneous function.

{\bf 3.6} It is well known that the conventional WDVV equations (\ref{FFFconv})
are pretty restrictive: this is an overdetermined system
of equations for a single function $F(t_i)$, and it is a kind
of surprise that they possess any solutions at all, and in fact there
exist vast variety of them (associated with Whitham
hierarchies, topological models and quantum cohomologies).
The set (\ref{FFF}) is even more overdetermined than
(\ref{FFFconv}), since $k$ can take {\it any} value. Thus, it is even
more surprising that the solutions still
exist (in order to convince the reader, we supplement the
formal proof in s.4 below by explicit
examples in Appendices A and B).

Of course, (\ref{FFF}) is tautologically true
for $N=2$ and $N=3$, it becomes a non-trivial system for
$N\geq 4$.

{\bf 3.7}  Our proof in s.4 actually suggests that
in majority of cases when the {\it ordinary} WDVV
(\ref{FFFconv}) is true, the whole system (\ref{FFF})
holds automatically. This implies that this entire
system should possess some interpretation in the
spirit of hierarchies or hidden symmetries. It still
remains to be found.
The geometrical or cohomological origin of relations (\ref{FFF})
also remains obscure.

{\bf 3.8}  In this paper we discuss solutions to (\ref{FFF}),
provided by conventional topological theories and
-- as a far less trivial example -- by the simplest
Seiberg-Witten prepotentials.

We beleive that more solutions to (\ref{FFF}) can arise from
more sophisticated examples of the Seiberg-Witten
theory (${\cal N}=2$ SUSY Yang-Mills with other groups and
with matter supermultiplets); the most interesting should be
the UV-finite models, when hyperelliptic surfaces
(the double coverings of $CP^1$) are
substituted by coverings of elliptic curve (torus), and  a new
elliptic parameter $\tau$ emerges.

If this conjecture is true, one can look for some relation
between (\ref{FFF}) and Picard-Fuchs equations, and then
address to the issue
of the WDVV equations for the prepotential, associated with
families of the Calabi-Yau manifolds.

{\bf 3.9} Effective theory (\ref{1}) is naively {\it non-\-topological}.
From the 4-dimensional point of view it describes the
low-energy limit of the Yang-Mills theory which -- at least, in
the ${\cal N}=2$ supersymmetric case -- is {\it not} topological and
contains propagating massless particles. Still this theory
is entirely defined by a prepotential, which -- as we now
see -- possesses {\it all} essential properties of the
prepotentials in topological theory. Thus, from the
``stringy'' point of view (when everything is described
in terms of universality classes of effective actions)
the Seiberg-Witten models belong to the same class as
topological models: only the way to extract physically
meaningful correlators from the prepotential is
different. This can serve as a new evidence
that the notion of topological theories is deeper than
it is usually assumed: as emphasized in \cite{GKMMM} it
can be actually more related to the low-energy (IR) limit of
field theories than to the property of the correlation
functions to be constants in physical space-time.

{\bf 3.10} The issue of the WDVV equations in context of the
Seiberg-Witten theory  has been addressed
in \cite{BM}. Unfortunately, we do
not understand the statements in this paper and their relation to
eqs.(\ref{FFF}).

\section{The proof of eqs.(5)}

{\bf 4.1} Let us begin with reminding the proof of the WDVV equations
(\ref{FFFconv}) for ordinary topological theories.
We take the simplest of all possible examples, when
$\phi_i$ are polynomials of a single variable $\lambda$.
The proof is essentially the check of consistency between the
following formulas:
\be
\phi_i(\lambda)\phi_j(\lambda) = C_{ij}^k\phi_k(\lambda)
\ {\rm mod}\ W'(\lambda),
\label{.c}
\ee
\be
F_{ijk} = {\rm res}\frac{\phi_i\phi_j\phi_k(\lambda)}
{W'(\lambda)} = \sum_{\alpha}
\frac{\phi_i\phi_j\phi_k(\lambda_\alpha)}
{W''(\lambda_\alpha)},
\label{vc}
\ee
\be
\eta_{kl} = {\rm res}\frac{\phi_k\phi_l(\lambda)}
{W'(\lambda)} = \sum_{\alpha}
\frac{\phi_k\phi_l(\lambda_\alpha)}
{W''(\lambda_\alpha)},
\label{vvc}
\ee
\be
F_{ijk} = \eta_{kl}C_{ij}^l.
\label{vvvc}
\ee
Here $\lambda_\alpha$ are the roots of $W'(\lambda)$.

In addition to the consistency of (\ref{.c})-(\ref{vvvc}),
one should know that {\it such} $F_{ijk}$, given by
(\ref{vc}), are the third derivatives of a single function $F(a)$, i.e.
\be
F_{ijk} = \frac{\partial^3F}{\partial a_i
\partial a_j\partial a_k}.
\ee
This integrability property of (\ref{vc}) follows from separate
arguments and can be checked independently.
But if (\ref{.c})-(\ref{vvc}) is given, the proof of
(\ref{vvvc}) is straightforward:
\be \eta_{kl}C^l_{ij} = \sum_{\alpha}
\frac{\phi_k\phi_l(\lambda_\alpha)}
{W''(\lambda_\alpha)} C^l_{ij} \stackrel{(\ref{.c})}{=} \\ =
\sum_{\alpha}
\frac{\phi_k(\lambda_\alpha)}
{W''(\lambda_\alpha)} \phi_i(\lambda_\alpha)\phi_j(\lambda_\alpha)
= F_{ijk}.
\ee
Note that (\ref{.c}) is defined modulo $W'(\lambda)$,
but $W'(\lambda_\alpha) = 0$ at all the points $\lambda_\alpha$.

Imagine now that we change the definition of the metric:
\be
\eta_{kl} \rightarrow \eta_{kl}(\omega) =
\sum_{\alpha}
\frac{\phi_k\phi_l(\lambda_\alpha)}
{W''(\lambda_\alpha)}\omega(\lambda_\alpha).
\ee
Then the WDVV equations would still be correct, provided the
definition (\ref{.c}) of the algebra is also changed for
\be
\phi_i(\lambda)\phi_j(\lambda) = C_{ij}^k(\omega)\phi_k(\lambda)
\omega(\lambda)\ {\rm mod}\ W'(\lambda).
\label{..c}
\ee
This describes an associative algebra, whenever the
polynomials $\omega(\lambda)$ and $W'(\lambda)$ are co-prime,
i.e. do not have common divisors.
Note that (\ref{vc}) -- and thus the fact that $F_{ijk}$
is the third derivative of the same $F$ -- remains intact!
One can now take for $\omega(\lambda)$ any of the operators
$\phi_k(\lambda)$, thus reproducing eqs.(\ref{FFF}) for
all topological theories
\footnote{To make (\ref{FFF})
sensible, one should require that $W'(\lambda)$ has only
{\it simple} zeroes, otherwise some of the matrices $F_k$
can be degenerate and non-invertible.
}
(see Appendix A for explicit example).

{\bf 4.2} In the case of the Seiberg-Witten model the polynomials
$\phi_i(\lambda)$ are substituted by the canonical holomorphic
differentials $d\omega_i(\lambda )$ on hyperelliptic surface
(\ref{hesur}). This surface
can be represented in a standard hyperelliptic form,
\be
y^2 = P_N^2(\lambda ) - 1,
\ee
(where $y = \frac{1}{2}\left(w - \frac{1}{w}\right)$) and is
of genus $g = N-1$.\footnote{
Note that in this way one defines a peculiar $g$-parametric
family of hyperelliptic surfaces (the moduli space of {\it all}
the Riemann surfaces has dimension $3g-3$, while that of all the
hyperelliptic ones -- $2g-1$). One can take for the $g$ moduli
the set $\{h_k\}$ or instead the set of periods $\{a_i\}$.
This particular family is associated with the Toda-chain
hierarchy, $N$ being the length of the chain (while {\it all}
the Riemann surfaces of all genera are associated with KP,
and {\it all} the hyperelliptic ones -- with KdV hierarchy).
}

{\bf 4.2.1} Instead of (\ref{.c}) and (\ref{..c}) we now put
\be
d\omega_i(\lambda )d\omega_j(\lambda ) =
C_{ij}^k(d\omega) d\omega_k(\lambda )
d\omega(\lambda ) \ {\rm mod}\ \frac{dP_N(\lambda )d\lambda }{y^2}.
\label{.}
\ee
In contrast to (\ref{..c}) we can not now choose $\omega = 1$
(to reproduce (\ref{.c})), because now we need it to be
a 1-differential. Instead we just take $d\omega$ to be a
{\it holomorphic} 1-differential. However, there is no distinguished
one -- just a $g$-parametric family -- and $d\omega$ can be
{\it any} one from this family. We require only that it is
co-prime with $\frac{dP_N(\lambda )}{y}$.

If the algebra (\ref{.}) exists, the structure constants
$C_{ij}^k(d\omega)$ satisfy the associativity condition
(if $d\omega$ and
${dP_N\over y}$ are co-prime). But we still need to show that
it indeed exists, i.e. that if $d\omega$ is given, one can find
($\lambda $-independent) $C_{ij}^k$. This is a simple exercise:
all $d\omega_i$ are linear combinations of
\be
dv_k(\lambda ) = \frac{\lambda ^{k-1}d\lambda }{y}, \ \ \ k=1,\ldots,g: \\
dv_k(\lambda ) = \sigma_{ki}d\omega_i(\lambda ), \ \ \
d\omega_i = (\sigma^{-1})_{ik}dv_k, \ \ \
\sigma_{ki} = \oint_{A_i}dv_k,
\label{sigmadef}
\ee
also $d\omega(\lambda ) = s_kdv_k(\lambda )$.
Thus, (\ref{.}) is in fact a relation between the polynomials:
\be
\left(\sigma^{-1}_{ii'}\lambda ^{i'-1}\right)
\left( \sigma^{-1}_{jj'}\lambda ^{j'-1}\right) =
C_{ij}^k \left(\sigma^{-1}_{kk'}\lambda ^{k'-1}\right)
\left( s_l\lambda ^{l-1}\right) +
p_{ij}(\lambda )P'_N(\lambda ).
\ee
At the l.h.s. we have a polynomial of degree $2(g-1)$.
Since $P'_N(\lambda )$ is a polynomial of degree $N-1=g$, this
implies that $p_{ij}(\lambda )$ should be a polynomial of degree
$2(g-1)-g = g-2$. The identification of two polynomials of
degree $2(g-1)$ impose a set of $2g-1$ equations for the coefficients.
We have a freedom to adjust $C_{ij}^k$ and $p_{ij}(\lambda )$
(with $i,j$ fixed), i.e. $g + (g-1) = 2g-1$ free parameters:
exactly what is necessary. The linear system of equations
is non-degenerate for co-prime $d\omega$ and $dP_N/y$.

Thus, we proved that the algebra (\ref{.}) exists (for a given
$d\omega$) -- and thus $C_{ij}^k(d\omega)$ satisfy the
associativity condition
\be
C_i(d\omega)C_j(d\omega) = C_j(d\omega) C_i(d\omega).
\ee

{\bf 4.2.2} Instead of (\ref{vc}) we have \cite{KDM}:
\be
F_{ijk} = \frac{\partial^3F}{\partial a_i\partial a_j
\partial a_k} = \frac{\partial T_{ij}}{\partial a_k} = \nn \\
= \stackreb{d\lambda =0}{{\rm res}} \frac{d\omega_id\omega_j
d\omega_k}{d\lambda\left(\frac{dw}{w}\right)} =
\stackreb{d\lambda =0}{{\rm res}} \frac{d\omega_id\omega_j
d\omega_k}{d\lambda\frac{dP_N}{y}} =
\sum_{\alpha} \frac{\hat\omega_i(\lambda_\alpha)\hat\omega_j
(\lambda_\alpha)\hat\omega_k(\lambda_\alpha)}{P'_N(\lambda_\alpha)
/\hat y(\lambda_\alpha)}
\label{v}
\ee
The sum at the r.h.s. goes over all the $2g+2$ ramification points
$\lambda_\alpha$ of the hyperelliptic curve (i.e. over the zeroes
of $y^2 = P_N^2(\lambda )-1 =
\prod_{\alpha=1}^N(\lambda - \lambda_\alpha)$);
\  $d\omega_i(\lambda) = (\hat\omega_i(\lambda_\alpha) +
O(\lambda-\lambda_\alpha))\frac{d\lambda}
{\sqrt{\lambda-\lambda_\alpha}}$,
\ $\ \ \ \hat y^2(\lambda_\alpha) =
\prod_{\beta\neq\alpha}(\lambda_\alpha - \lambda_\beta)$.

Though eq.(\ref{v}) can be extracted from \cite{KDM},
for the sake of completeness we present a proof of this formula
in Appendix C at the end of this paper.

{\bf 4.2.3} We define the metric in the following way:
\be
\eta_{kl}(d\omega) =
\stackreb{d\lambda =0}{{\rm res}} \frac{d\omega_kd\omega_l
d\omega}{d\lambda\left(\frac{dw}{w}\right)} =
\stackreb{d\lambda =0}{{\rm res}} \frac{d\omega_kd\omega_l
d\omega_k}{d\lambda\frac{dP_N}{y}} = \\ =
\sum_{\alpha} \frac{\hat\omega_k(\lambda_\alpha)\hat\omega_l
(\lambda_\alpha)\hat\omega(\lambda_\alpha)}{P'_N(\lambda_\alpha)
/\hat y(\lambda_\alpha)}
\label{vv}
\ee
In particular, for $d\omega = d\omega_k$,
$\eta_{ij}(d\omega_k) = F_{ijk}$: this choice will
give rise to (\ref{FFF}).

Given (\ref{.}), (\ref{v}) and (\ref{vv}), one can check:
\be
F_{ijk} = \eta_{kl}(d\omega)C_{ij}^k(d\omega).
\label{vvv}
\ee
Note that $F_{ijk} = {\partial^3F\over\partial a_i\partial a_j
\partial a_k}$ at the l.h.s. of (\ref{vvv}) is independent
of $d\omega$! The r.h.s. of (\ref{vvv}) is equal to:
\be
\eta_{kl}(d\omega)C_{ij}^k(d\omega) =
\stackreb{d\lambda =0}{{\rm res}} \frac{d\omega_kd\omega_l
d\omega}{d\lambda\left(\frac{dw}{w}\right)} C_{ij}^l(d\omega)
\stackrel{(\ref{.})}{=} \\ =
\stackreb{d\lambda =0}{{\rm res}} \frac{d\omega_k}
{d\lambda\left(\frac{dw}{w}\right)}
\left(d\omega_id\omega_j - p_{ij}\frac{dP_Nd\lambda}{y^2}\right) =
F_{ijk} - \stackreb{d\lambda =0}{{\rm res}} \frac{d\omega_k}
{d\lambda\left(\frac{dP_N}{y}\right)}p_{ij}(\lambda)\frac{dP_Nd\lambda}{y^2}
= \\ = F_{ijk} - \stackreb{d\lambda =0}{{\rm res}}
\frac{p_{ij}(\lambda )d\omega_k(\lambda)} {y}
\ee
It remains to prove
that the last item is indeed vanishing for any $i,j,k$.
This follows from the
fact that $\frac{p_{ij}(\lambda )d\omega_k(\lambda )}{y}$
is singular only at zeroes of $y$, it is not singular at
$\lambda =\infty$ because
$p_{ij}(\lambda)$ is a polynomial of low enough degree
$g-2 < g+1$. Thus the sum of its residues at ramification points
is thus the sum over {\it all} the residues and therefore vanishes.

This completes the proof of associativity condition for any $d\omega$.
Taking $d\omega = d\omega_k$ (which is obviously co-\-prime with
$\frac{dP_N}{y}$), we obtain (\ref{FFF}).

\section{Appendix A. Explicit example of (5) for
topological theory}

In this appendix we address to the questions about the
system (\ref{FFF}) with the two goals:
First, we provide explicit examples to convince
the reader that entire system (\ref{FFF}) is generically
true for topological theories, not only (\ref{FFFconv}),
as one usually believes.
Second -- since one gets convinced --
we ask if (\ref{FFF}) is just a direct corollary of
(\ref{FFFconv}), supplemented by peculiar symmetry
properties $(F_i)_{jk} = (F_j)_{ik} = (F_k)_{ij}$.
We demonstrate that this is indeed the case for $g = N-1 = 3$
(eqs.(\ref{FFF}) are tautologically correct for $g=1$
and $g=2$). However, this does not seem to be the case for
$g\geq 4$: (\ref{FFF}) relies heavily
on the fact that $F_{ijk}$ are the third derivatives,
$F_{ijk} = {\partial^3 F\over\partial t_i\partial t_j\partial t_k}$,
namely on relations like (\ref{vc}).

\subsection{Examples for $g = N-1 = 3$}

{\bf 5.1.1} Let us begin with the topological model with
$W'(\lambda) = \lambda^3 - q$ ($q\neq 0$ -- the roots of
$W'(\lambda)$ are all different -- in order to avoid degeneracies
of the matrices $F_1$ and $F_2$). In the basis
$\phi_i = \lambda^i$, $i=0,1,2$ one easily obtains from
(\ref{vc}):
\be
F_0 = \left(\begin{array}{ccc}
0&0&1\\0&1&0\\1&0&0 \end{array}\right), \ \ \ \
F_1 = \left(\begin{array}{ccc}
0&1&0\\1&0&0\\0&0&q\end{array}\right), \ \ \ \
F_2 = \left(\begin{array}{ccc}
1&0&0\\0&0&q\\0&q&0 \end{array}\right).
\label{Fexpl}
\ee
The corresponding prepotential is
\be
F = \frac{1}{2}t_0t_1^2 + \frac{1}{2}t_0^2t_2 +
\frac{q}{2}t_1t_2^2.
\ee
The inverse matrices are $F_i^{-1}(q) = F_i(1/q)$.

In order to shorten the calculations it is useful
to note that -- since the matrices $F_i$ are symmetric -- the
relations (\ref{FFF}) mean that all the matrices
\be
U_{ikj} = F_iF_k^{-1}F_j: \ \ \ \
U_{ikj} = U_{ikj}^{tr}.
\ee
are also symmetric.
It is a trivial exercise to check that
\be
U_{102} = F_1F_0^{-1}F_2 = qI, \ \ \
U_{201} = qF_1, \ \ \
U_{012} = F_1(q), \ \ \
U_{210} = F_1(1/q), \nn \\
U_{021} = U_{120} =
\left(\begin{array}{ccc}
1/q&0&0\\0&0&1\\0&1&0\end{array}\right)
\ee
are indeed all symmetric.

{\bf 5.1.2} Consider now a generalization of the
previous example: the quantum cohomology
of $CP^2$ \cite{KM}. The prepotential is
\be
F = \frac{1}{2}t_0t_1^2 + \frac{1}{2}t_0^2t_2 +
\sum_{n=1}^\infty
\frac{N_n t_2^{3n-1}}{(3n-1)!}e^{nt_1}
\ee
and the corresponding matrices are:
\be
F_0 = \left(\begin{array}{ccc}
0&0&1\\0&1&0\\1&0&0\end{array}\right), \ \ \
F_1 = \left(\begin{array}{ccc}
0&1&0\\1&F_{111}&F_{112}\\0&F_{112}&F_{122}\end{array}\right), \ \ \
F_2 = \left(\begin{array}{ccc}
1&0&0\\0&F_{112}&F_{122}\\0&F_{122}&F_{222}\end{array}\right)
\ee
where
\be
F_{111} = \sum_n \frac{n^3N_n}{(3n-1)!} t_2^{3n-1}e^{nt_1},\\
F_{112} = \sum_n \frac{n^2N_n}{(3n-2)!} t_2^{3n-2}e^{nt_1},\\
F_{122} = \sum_n \frac{nN_n}{(3n-3)!} t_2^{3n-3} e^{nt_1}, \\
F_{222} = \sum_n \frac{N_n}{(3n-4)!} t_2^{3n-4} e^{nt_1}.
\ee
One can easily check that every equation in (\ref{FFF})
is true if and only if
\be
F_{222} = F_{112}^2 - F_{111}F_{122}.
\label{ququ}
\ee
Indeed,
\be
F_1F_0^{-1}F_2 = \left(\begin{array}{ccc}
0 & F_{112} & F_{122} \\
F_{112} & F_{122} + F_{111}F_{112} & F_{222}+F_{111}F_{122} \\
F_{122} & F_{112}^2 & F_{112}F_{122} \end{array}\right), \nn \\
F_0F_1^{-1}F_2 = \frac{1}{F_{122}}\left(\begin{array}{ccc}
-F_{112} & F_{122} & F_{222} \\
F_{122} & 0 & 0 \\
F_{112}^2-F_{111}F_{122} & 0 & F_{122}^2-F_{112}F_{222}
\end{array}\right), \nn \\
F_0F_2^{-1}F_1 = \frac{1}{F_{112}F_{222}-F_{122}^2}
\left(\begin{array}{ccc}
-F_{122} & F_{112}^2-F_{111}F_{122} & 0 \\
F_{222} & F_{111}F_{222}-F_{112}F_{122} &
F_{112}F_{222}-F_{122}^2 \\
0 & F_{112}F_{222}-F_{122}^2 & 0 \end{array}\right)
\ee
Eq.(\ref{ququ}) is the famous equation,
providing the recursive relations for $N_n$ \cite{KM}:
\be \frac{N_n}{(3n-4)!} = \sum_{a+b=n}
\frac{a^2b(3b-1)b(2a-b)}{(3a-1)!(3b-1)!}N_aN_b.
\ee
For example, $N_2=N_1^2$, $N_3 = 12N_1N_2 = 12N_1^3$, $\ldots$

The zero curvature condition (\ref{FFFD}) is obviously
satisfied for $\eta^{0} = F_0$: $R_{ijkl}(\eta^{(0)}) = 0$, but
it is not fulfilled already for $\eta^{(1)} = F_1$:
\be
R_{1212}(\eta^{(1)}) \sim F_{1112}F_{1222} - F_{1122}^2 =
-N_1^2 e^{3t_1} + \ldots \neq 0.
\ee

{\bf 5.1.3} Two above examples illustrate that -- if
(\ref{FFFconv}), i.e. relation for $k=0$, is established --
the equations (\ref{FFF}) for all other $k$ hold as well. This of course
follows -- for the topological systems -- from our analysis
in s.4.1, but in fact for $g = N-1 = 3$ this is just
an {\it arithmetic} property: one should only take into
account the fact that $F_{ijk}$ is symmetric in all three
indices.

Namely, let us write down the only non-trivial matrix
element in relation (\ref{FFFconv}):
\be
(F_{11i}F_{22j} - F_{12i}F_{12j})(F_0^{-1})^{ij} = 0,
\label{*}
\ee
$(F_0^{-1})^{ij} = (\det\ F_0)^{-1} \hat F_0^{ij}$,
where the entries in $\hat F_0$ are quadratic combinations
of $F_{klm}$. Substituting the explicit expression for
$\hat F_0$, we get for (\ref{*}) certain sophisticated
expression (too long to be presented here)
through the 4-th powers of $F_{klm}$.

Now, do the same for the other eqs. in (\ref{FFF}),
e.g. for $U_{012}$: the only non-trivial matrix element
is
\be
(F_{00i}F_{22j} - F_{02i}F_{02j})(F_1^{-1})^{ij} =
(\det\ F_1)^{-1}\times({\rm quartic\ combination\ of}\
F_{klm}).
\label{**}
\ee
One can check that the quartic combinations are literally the
same in (\ref{*}) and (\ref{**}) -- and in all other $U_{ijk}$,
i.e. if any one of the equations (\ref{FFF}) is satisfied,
the others follow arithmetically.

\subsection{$g > 3$}

Thus, we see that for $g = N-1 = 3$ any solution to the original
WDVV eq. (\ref{FFFconv}) is just {\it literally} solution to the
whole system (\ref{FFF}).

We now argue that for $g\geq 4$ this is -- though generically true --
but not for such  a simple reason.
Then we provide an analogue of the example from
s.5.1.1 for $g\geq 4$ -- which is now a little less trivial
illustration.

{\bf 5.2.1} Let us try to repeat the reasoning from s.5.1.3
for generic $g$. The matrix element
\be
\left(F_iF_k^{-1}F_j - F_jF_k^{-1}F_i\right)_{mn}
= (F_{imr}F_{jns} - F_{inr}F_{jms})(F_k^{-1})^{rs} = \nn \\ =
(\det\ F_k)^{-1}\epsilon^{rr_1\ldots r_{g-1}}
\epsilon^{ss_1\ldots s_{g-1}}
(F_{imr}F_{jns} - F_{inr}F_{jms})
F_{kr_1s_1}\ldots F_{kr_{g-1}s_{g-1}}.
\label{b}
\ee

If $k=0$, but $i,j,m,n\neq 0$, the r.h.s. of (\ref{b})
contains exactly $g+1$ indices "0" ($g-1$ times $k=0$
plus exactly one of all the $r$'s and exactly one of all the
$s$'s). Of indices $i,j,m,n$ at most two can be equal
to 0 without making (\ref{b}) vanishing identically.
Thus, every item at the r.h.s. of (\ref{b}) for $k=0$
contains $g+1$, $g+2$ or $g+3$ indices "0".

If $k\neq 0$, and $i,j,m,n\neq 0$, the number of indices
"0" at the r.h.s. is exactly 2 (one of all the $r$'s and
one of all the $s$'s). Adding at most 2 indices "0"
from among $i,j,m,n$ we get 2,3 or 4 such indices in every
item if $k\neq 0$.

If entire system (\ref{FFF}) with all $k$'s
was {\it arithmetic} corollary
of its subset (\ref{FFFconv}) with $k=0$ -- as is the case
for $g=3$ in s.5.1.3 -- the number of all indices, including
"0",  should match, i.e. $g+1$, $g+2$ or $g+3$ should
coincide with 2,3 or 4. This restricts $g$ to be $g\leq 3$.
For $g\geq 4$ the implication (\ref{FFFconv})
$\Longrightarrow$ (\ref{FFF}) -- still true according to
our consideration in s.4 -- should be of more transcendental
nature.

{\bf 5.2.2} Now we take the topological theory with $W'(\lambda) =
\lambda^g - q$. In the basis $\phi_i = \lambda^i$,
$i = 0,\ldots,g-1$ matrices $F_i$ are $g\times g$
analogs of (\ref{Fexpl}), now units stand at the
$i$-th upper skew-\-subdiagonal and $q$'s -- at the
$(q+1-i)$-th lower one so, again,
$F_i^{-1}(q) = F_i(1/q)$: this is enough for explicit
calculation.

For example, for $g=4$ not only conventional combinations
$U_{i0j} = F_iF_0^{-1}F_j$ are symmetric (i.e. satisfy
(\ref{FFFconv}), e.g.
\be
U_{102} = \left(\begin{array}{cccc}
   0&0&1&0\\ 0&1&0&0 \\ 1&0&0&0 \\ 0&0&0&q
           \end{array}\right)
\left(\begin{array}{cccc}
   0&0&0&1\\ 0&0&1&0 \\ 0&1&0&0 \\ 1&0&0&0
           \end{array}\right)
\left(\begin{array}{cccc}
   0&1&0&0\\ 1&0&0&0 \\ 0&0&0&q \\ 0&0&q&0
           \end{array}\right)  =
\left(\begin{array}{cccc}
   1&0&0&0\\ 0&0&0&q \\ 0&0&q&0 \\ 0&q&0&0
           \end{array}\right),
\ee
but the same is true, say, for
\be
U_{123} = F_1F_2^{-1}F_3 = \left(\begin{array}{cccc}
   0&0&1&0\\ 0&1&0&0 \\ 1&0&0&0 \\ 0&0&0&q
           \end{array}\right)
\left(\begin{array}{cccc}
   0&1&0&0\\ 1&0&0&0 \\ 0&0&0&1/q \\ 0&0&1/q&0
           \end{array}\right)
\left(\begin{array}{cccc}
   1&0&0&0\\ 0&0&0&q \\ 0&0&q&0 \\ 0&q&0&0
           \end{array}\right)  =
\nn \\
= \left(\begin{array}{cccc}
   0&1&0&0\\ 1&0&0&0 \\ 0&0&0&q \\ 0&0&q&0
           \end{array}\right)
\ee
and all other $U_{ijk}$.

\section{Appendix B. Explicit example of (5) related to
the Seiberg-Witten effective theory}

This example involves the leading (perturbative) approximation
to the exact Seiberg-Witten prepotential,  which -- being the
leading contribution -- satisfies (\ref{FFF}) by itself.
The perturbative contribution is non-transcendental, thus
calculation can be performed in explicit form:
\be
F_{pert} \equiv F(a_i) =
\left.\frac{1}{2}\sum_{\stackrel{m<n}{m,n=1}}^N
(A_m-A_n)^2\log(A_m-A_n)\right|_{\sum_m A_m = 0} = \nn \\
= \frac{1}{2}\sum_{\stackrel{i<j}{i,j=1}}^{N-1}
(a_i-a_j)^2\log(a_i-a_j) +
\frac{1}{2}\sum_{i=1}^{N-1}a_i^2\log a_i
\label{pertF}
\ee
Here we took $a_i = A_i - A_N$ -- one of the many
possible choices of independent variables, which differ by
linear transformations. According to comment {\bf 3.3}
above, the system (\ref{FFF}) is covariant under such changes.

We shall use the notation $a_{ij} = a_i - a_j$. The matrix
\be
\{(F_1)_{mn}\} = \left\{\frac{\partial^3 F}{\partial a_1
\partial a_m\partial a_n} \right\} = \nn \\ =
\left(\begin{array}{ccccc}
\frac{1}{a_1} +\sum_{l\neq 1} \frac{1}{a_{1l}} &
-\frac{1}{a_{12}} & -\frac{1}{a_{13}} & -\frac{1}{a_{14}} & \\
-\frac{1}{a_{12}}& +\frac{1}{a_{12}} & 0 & 0 & \\
-\frac{1}{a_{13}}& 0 & +\frac{1}{a_{13}}& 0 &\ldots \\
-\frac{1}{a_{14}}& 0 & 0 &+\frac{1}{a_{14}} & \\
&&\ldots && \end{array}\right)
\ee
i.e.,
\be\label{f}
\{(F_i)_{mn}\} = \frac{\delta_{mn}(1-\delta_{mi})(1-\delta_{ni})}
{a_{im}} - \frac{\delta_{mi}(1-\delta_{ni})}{a_{in}}
- \frac{\delta_{ni}(1-\delta_{mi})}{a_{im}} + \nn \\ +
\left(\frac{1}{a_i} + \sum_{l\neq i}\frac{1}{a_{ik}}\right)
\delta_{mi}\delta_{ni}
\ee
The inverse matrix
\be\label{f-1}
\{(F_k^{-1})_{mn}\} = a_k + \delta_{mn}a_{km}(1-\delta_{mk}),
\ee
for example
\be
\{(F_1^{-1})_{mn}\} = a_1\left(\begin{array}{cccc}
1 & 1 & 1 & . \\ 1 & 1 & 1 & . \\ 1 & 1 & 1 & . \\
&\ldots & & \end{array}\right) +
\left(\begin{array}{cccc}
0 & 0 & 0 & . \\
0 & a_{12} & 0 & . \\
0 & 0 & a_{13} & . \\
& \ldots & & \end{array}\right)
\ee
As the simplest example let us consider the case $N=4$.
We already know from s.5.1.3 that for $N=4$ it is
sufficient to check only one of the eqs.(\ref{FFF}),
all the others follow automatically. We take $k=1$. Then,
\be
F_1=\left(
\begin{array}{ccc}
{1\over a_1}+{1\over a_{12}}+{1\over a_{13}}&-{1\over a_{12}}&-{1\over
a_{13}} \\-{1\over a_{12}}&{1\over a_{12}}&0\\-{1\over a_{13}}&0&{1\over
a_{13}}
\end{array}\right)\;,\ \ F^{-1}_2=\left(
\begin{array}{ccc}
a_2+a_{21}&a_2&a_2\\a_2&a_2&a_2\\a_2&a_2&a_2+a_{23}
\end{array}\right)\;,\\F_3=\left(
\begin{array}{ccc}
{1\over a_{31}}&0&-{1\over a_{31}}\\
0&{1\over a_{32}}&-{1\over a_{32}}\\
-{1\over a_{31}}&-{1\over a_{32}}&{1\over a_3}+{1\over a_{31}}+{1\over
a_{32}}
\end{array}\right)
\ee
and, say,
\be
F_1F^{-1}_2F_3=\left(
\begin{array}{ccc}
\star&-{1\over a_{31}}& \Delta + {a_{21}+a_{23}\over a_{13}^2}\\
-{1\over a_{13}}&\star&{1\over a_{13}}\\
{a_{21}+a_{23}\over a_{13}^2}&{1\over a_{13}}&\star
\end{array}\right)
\ee
where we do not write down manifestly the diagonal terms since, to check
(\ref{FFF}), one only needs to prove the symmetricity of the matrix. This is
really the case, since
\be
\Delta\equiv
{a_2\over a_1a_3}-{a_{21}\over a_1a_{31}}-{a_{23}\over a_3a_{13}} =0
\ee
Only at this stage we use manifestly that $a_{ij}=a_i-a_j$.

Now let us prove (\ref{FFF}) for the general case. We check the equation
for the inverse matrices. Namely, using formulas (\ref{f})-(\ref{f-1}), one
obtains
\be\label{long}
(F_i^{-1}F_jF_k^{-1})_{\alpha\beta}={a_ia_k\over a_j}+
\delta_{\alpha\beta}(1-\delta_{i\alpha})(1-\delta_{k\alpha})
(1-\delta_{j\alpha}){a_{i\alpha}a_{k\beta}\over a_{j\beta}}+
\delta_{j\alpha}\delta_{j\beta}(1-\delta_{i\alpha})(1-\delta_{k\beta})
\left({1\over a_j}+\sum_{n\ne j}{1\over a_{jn}}\right)+\\
+\delta_{j\alpha}(1-\delta_{i\alpha})a_{i\alpha}\left(
{a_k\over a_j}-{a_{k\beta}\over a_{j\beta}}(1-\delta_{k\beta})
(1-\delta_{j\beta})\right)+\delta_{j\beta}(1-\delta_{k\beta})\left(
{a_i\over a_j}-{a_{i\alpha}\over a_{j\alpha}}(1-\delta_{i\alpha})
(1-\delta_{j\alpha})\right)=\\
={a_ia_k\over a_j}+
\delta_{\alpha\beta}(1-\delta_{i\alpha}-\delta_{k\alpha}
-\delta_{j\alpha}){a_{i\alpha}a_{k\beta}\over a_{j\beta}}+
\delta_{j\alpha}\delta_{j\beta}\left({1\over a_j}+
\sum_{n\ne j}{1\over a_{jn}}\right)+\\
+\delta_{j\alpha}a_{i\alpha}\left(
{a_k\over a_j}-{a_{k\beta}\over a_{j\beta}}(1-\delta_{k\beta}
-\delta_{j\beta})\right)+\delta_{j\beta}\left(
{a_i\over a_j}-{a_{i\alpha}\over a_{j\alpha}}(1-\delta_{i\alpha}
-\delta_{j\alpha})\right)
\ee
where we used that $i\ne j\ne k$. The first three terms are evidently
symmetric with respect to interchanging $\alpha\leftrightarrow\beta$. In
order to prove the symmetricity of the last two terms, we need to use the
identities ${a_k\over a_j}-{a_{k\beta}\over a_{j\beta}}={a_{\beta}a_{jk}\over
a_ja_{j\beta}}\stackrel{k=\beta}{\to}{a_k\over a_j}$,
${a_i\over a_j}-{a_{i\alpha}\over a_{j\alpha}}=
{a_{\alpha}a_{ji}\over a_ja_{j\alpha}}\stackrel{i=\alpha}{\to}{a_i\over
a_j}$. Then, one gets
\be
\hbox{the last line of (\ref{long})}=
\delta_{j\alpha}(1-\delta_{j\beta})
{a_{ij}a_{jk}\over a_j}{a_{\beta}\over a_{j\beta}} +
\delta_{j\beta}(1-\delta_{j\alpha})
{a_{ij}a_{jk}\over a_j}{a_{\alpha}\over a_{j\alpha}} +
\delta_{j\alpha}\delta_{j\beta}{a_ka_{i\alpha}+a_ia_{k\beta}\over a_j}
\ee

It is interesting to note (see also comment {\bf 3.4}) that
in the particular example (\ref{pertF}),
all the metrics $\eta^{(k)}$ are flat. Moreover,
it is easy to find the explicit flat coordinates:
\be
\eta^{(k)} = \eta^{(k)}_{ij}da^ida^j =
F_{ijk}da^ida_j = da_ida_j\partial^2_{ij}(\partial_k F) = \nn \\ =
\frac{da_k^2}{a_k} + \sum_{l\neq k}\frac{da_{kl}^2}{a_{kl}} = 4\left(
(d\sqrt{a_k})^2 + \sum_{l\neq k}(d\sqrt{a_{kl}})\right).
\ee

\section{Appendix C. The proof of eq.(23)}

The crucial property of the differential  $dS$
is that its variation with respect to moduli is holomorphic 1-differential:
$\delta dS \cong \ holomorphic$, in fact
$\frac{\partial dS}{\partial a_i} \cong d\omega_i$.

From (\ref{defF}) it follows now \cite{SW} that the second
derivative of the prepotential is period matrix of the
Riemann surface:
\be\label{T}
\frac{\partial^2 F}{\partial a_i\partial a_j} =
\oint_{B_i} \frac{\partial dS}{\partial a_j} =
\oint_{B_i} d\omega_j = T_{ij}.
\ee
Thus, the third derivative
\be
\frac{\partial^3 F}{\partial a_i\partial a_j\partial a_k} =
\frac{\partial T_{ij}}{\partial a_k}.
\label{derT}
\ee
It is very easy to evaluate the derivative of the period
matrix of hyperelliptic curve w.r.t. the variation
of any ramification point $\lambda_\alpha$ \cite{Fay}\footnote{
Indeed, from (\ref{sigmadef}) and (\ref{T}), one obtains
$$
{\partial T_{ij}\over \partial\lambda_{\alpha}}=\sigma^{-1}_{ik}
\sigma^{-1}_{jm}\left(\oint_{B_l}{\partial
v_k\over\partial\lambda_{\alpha}}\oint_{A_l}v_m-\oint_{A_l}{\partial
v_k\over\partial\lambda_{\alpha}}\oint_{B_l}v_m\right)
$$
Using the local representation $v_m=du_m$, one gets
$$
0=\int v_m\wedge {\partial v_k\over\partial\lambda_{\alpha}}=\int d\left(
u_m{\partial v_k\over\partial\lambda_{\alpha}}\right)=
\oint_{B_l}{\partial
v_k\over\partial\lambda_{\alpha}}\oint_{A_l}v_m-\oint_{A_l}{\partial
v_k\over\partial\lambda_{\alpha}}\oint_{B_l}v_m -
\stackreb{\lambda_{\alpha}}{\hbox{res}}\left(u_m
{\partial v_k\over\partial\lambda_{\alpha}}\right)
$$
Therefore,
$$
{\partial T_{ij}\over \partial\lambda_{\alpha}}=
\sigma^{-1}_{ik}\hat v_k(\lambda_{\alpha}) \sigma^{-1}_{jm}\hat v_m(
\lambda_{\alpha})=\hat\omega_i(\lambda_{\alpha})\hat\omega_j(\lambda_{\alpha})
$$
where we used the expansion in the vicinity of the point $\lambda_{\alpha}$:
$u_m=2\hat v_m(\lambda_{\alpha})\sqrt{\lambda-\lambda_{\alpha}} + \ldots$,
${\partial v_k\over\partial\lambda_{\alpha}}={\hat
v_k(\lambda_{\alpha})\over \lambda-\lambda_{\alpha}} + \ldots$.}:
\be
\frac{\partial
T_{ij}}{\partial \lambda_\alpha} = \hat
\omega_i(\lambda_\alpha)\hat\omega_j(\lambda_\alpha).
\label{derram}
\ee
However, for the family (\ref{hesur}) all the $2g+2$
ramification points depend only on $g$ moduli, thus we
should also know $\frac{\partial\lambda_\alpha}{\partial a_k}$.
This is easy to evaluate in two steps:
\be
\frac{\partial\lambda_\alpha}{\partial a_k} =
\frac{\partial\lambda_\alpha}{\partial h_l}
\frac{\partial h_l}{\partial a_k}.
\ee
First step: the derivative
\be
\frac{\partial a_k}{\partial h_l} = \oint_{A_k}
\frac{\partial dS}{\partial h_l}
\ee
can be found from the explicit expression for
$dS = \lambda\frac{dw}{w}$:
\be
\frac{\partial dS}{\partial h_l}
= {\rm exact\ form}\ - \frac{d\lambda}{w}\frac{\partial w}{\partial h_l}
= {\rm exact\ form}\ - \frac{\lambda^{l-1}d\lambda}{y}
\ee
(since $\frac{dw}{w} = \frac{dP_N}{y}$ and
$\frac{\partial P_N}{\partial h_l} = \lambda ^{l-1}$).
Thus
\be
\frac{\partial a_k}{\partial h_l} = -\oint_{A_k}dv_l
\stackrel{(\ref{sigmadef})}{=} - \sigma_{lk},
\ee
and
\be
\frac{\partial h_l}{\partial a_k} = -\sigma^{-1}_{kl}.
\ee
Second step: in order to evaluate $\frac{\partial\lambda_\alpha}
{\partial h_l}$, let us take $h_l$-derivative of
$P_N(\lambda ) = \prod_\beta (\lambda - \lambda_\beta)$ and then put
$\lambda = \lambda_\alpha$. We get first
\be
\lambda^{l-1} = -P_N(\lambda )\sum_\beta \frac{\partial\lambda_\beta}
{\partial h_l}\frac{1}{\lambda-\lambda_\beta},
\ee
and the only term in the sum at the r.h.s. which
contributes when $\lambda=\lambda_\alpha$ and $P_N(\lambda_\alpha) = 0$
is that with $\beta=\alpha$. Applying the L'H\^opital rule, we
obtain:
\be
\lambda_\alpha^{l-1} = -\frac{\partial \lambda_\alpha}{\partial h_l}
P_N'(\lambda_\alpha),
\ee
or
\be
\frac{\partial \lambda_\alpha}{\partial a_k} =
\frac{\lambda_\alpha^{l-1}}{P_N'(\lambda_\alpha)}
\sigma^{-1}_{kl} \stackrel{(\ref{sigmadef})}{=}
\frac{\hat y(\lambda_\alpha)}{P_N'(\lambda_\alpha)}
\hat\omega_k(\lambda_\alpha).
\ee
Together with (\ref{derT}) and (\ref{derram}) this
finally gives:
\be
\frac{\partial^3 F}{\partial a_i\partial a_j\partial a_k} =
\frac{\partial T_{ij}}{\partial a_k}
= \sum_\alpha \frac{\partial T_{ij}}{\partial\lambda_\alpha}
\frac{\partial \lambda_\alpha}{\partial a_k} = \nn \\ =
\sum_\alpha \hat\omega_i(\lambda_\alpha)
\hat\omega_j(\lambda_\alpha)
\frac{\hat\omega_k(\lambda_\alpha)}
{P_N'(\lambda_\alpha)/\hat y(\lambda_\alpha)}
\ee
as stated in (\ref{v}).

\section{Acknowledgements}

We are indebted to B.Dubrovin, A.Gorsky, S.Gukov, S.Kharchev,
I.Krichever, A.Losev, Yu.Manin, I.Polyubin and A.Rosly
for valuable discussions.
A.Mironov is grateful to the Institute of Theoretical Physics at Hannover
for the kind hospitality.
A.Morozov acknowledges the hospitality and support of the
Institute of Theoretical Physics at Helsinki
and Max-Planck Institute at Bonn, where parts of this work were done.
The work of A.Mar. is partially supported by grants INTAS-93-2058 and
RFFI-96-01-01106, the work of A.Mir. -- by grants INTAS-93-1038,
RFFI-96-02-16347a and Volkswagen Stiftung.

\end{document}